# A cosmic stream of atomic carbon gas connected to a massive radio galaxy at redshift 3.8

Bjorn H. C. Emonts[1,*], Matthew D. Lehnert[2,3], Ilsang Yoon[1], Nir Mandelker[4,5,6,7], Montserrat Villar-Martín[8], George K. Miley[9], Carlos De Breuck[10], Miguel A. Pérez-Torres[11,12,13], Nina A. Hatch[14], Pierre Guillard[3,15]

[1] National Radio Astronomy Observatory, Charlottesville, VA 22903, USA
[2] Université Lyon 1, École normale supérieure de Lyon, Centre National de la Recherche Scientifique, Unité mixte de recherche 5574, Centre de Recherche Astrophysique de Lyon, F-69230 Saint-Genis-Laval, France
[3] Sorbonne Université, Centre National de la Recherche Scientifique, Unité mixte de recherche 7095, Institut d'Astrophysique de Paris, 75014 Paris, France
[4] Racah Institute of Physics, The Hebrew University, Jerusalem 91904, Israel
[5] Kavli Institute for Theoretical Physics, University of California, Santa Barbara, CA 93106, USA
[6] Department of Astronomy, Yale University, New Haven, CT 06511, USA
[7] Heidelberger Institut für Theoretische Studien, Schloss-Wolfsbrunnenweg 35, D-69118 Heidelberg, Germany
[8] Centro de Astrobiología, Consejo Superior de Investigaciones Científicas – Instituto Nacional de Técnica Aeroespacial, 28850 Torrejón de Ardoz, Madrid, Spain
[9] Sterrewacht, Leiden University, 2300 RA Leiden, The Netherlands.
[10] European Southern Observatory, 85748 Garching, Germany
[11] Instituto de Astrofísica de Andalucía, Consejo Superior de Investigaciones Científicas, E-18008 Granada, Spain
[12] Departamento de Física Teórica, Facultad de Ciencias, Universidad de Zaragoza, 50019 Zaragoza, Spain.
[13] School of Sciences, European University Cyprus, Engomi, 1516 Nicosia, Cyprus
[14] School of Physics and Astronomy, University of Nottingham, Nottingham NG7 2RD, UK
[15] Institut Universitaire de France, Ministère de l'Enseignement Supérieur et de la Recherche, 75231 Paris, France

*Correspondence to: bemonts@nrao.edu

**The growth of galaxies in the early Universe is driven by accretion of circum- and inter-galactic gas. Simulations predict that steady streams of cold gas penetrate the dark matter halos of galaxies, providing the raw material necessary to sustain star formation. We report a filamentary stream of gas that extends for 100 kiloparsecs and connects to the massive radio galaxy 4C 41.17. The stream is detected using sub-millimeter observations of the [C I] line of atomic carbon, a tracer of neutral atomic or molecular hydrogen gas. The galaxy contains a central gas reservoir that is fueling a vigorous starburst. Our results show that the raw material for star formation can be present in cosmic streams outside galaxies.**

The growth and evolution of galaxies is thought to be linked to the accretion of gas from the circum- and inter-galactic medium. Simulations have shown that a large fraction of the gas that accretes into the dark matter halos that surround galaxies is not shock heated to high temperatures when passing through the accretion shock, but instead penetrates down into the halo as collimated steady flows of cold gas, referred to as accretion streams *(1, 2)*. Although this process could occur throughout the history of the Universe, the simulations predict it peaked from redshift 2 to 4 (roughly 10 to 12 billion years ago) *(3)*. Simulations that include these cold accretion streams match observations of high star formation rates in high-redshift galaxies *(4)*.

If the timescales for cooling of the accreting gas are shorter than the timescales for mixing of hot and cold gas, the streams are predicted to be stable against hydrodynamic instabilities, so additional gas from the circum-galactic medium cools and reinforces the streams *(5)*. Observations at optical wavelengths have detected accreting streams around distant massive galaxies, through the redshifted ultraviolet Lyman-α (Lyα) emission line of recombining hydrogen *(6-9)*. This Lyα emission has been interpreted as arising from hydrogen gas that is cooling as it flows towards the adjacent galaxy *(10-12)*. Although these results support the numerical simulations, Lyα emission only traces gas at temperatures of $\gtrsim 10^4$ K. Simulations lack the resolution to track the fragmentation that occurs when the gas cools to lower temperatures *(13)*. This limits predictions about accreting gas at temperatures of 10-100 K, at which point hydrogen is expected to be in the molecular phase. Molecular gas is the raw material that fuels star formation, so could potentially provide a direct link between cosmic gas accretion and galaxy growth.

## Observations of 4C 41.17

We used the Atacama Large Millimeter/submillimeter Array (ALMA) to search for cold gas in and around the massive galaxy 4C 41.17 (also known as B3 0647+415) *(14, 15)*, which is at redshift $z = 3.792$ *(16)*. 4C 41.17 consists of a group of small galaxies that are thought to be in the process of merging with a central radio galaxy, to form a single massive galaxy *(14, 15, 17)*. The central radio galaxy contains a supermassive black hole that generates jets visible at radio wavelengths *(14)*. The entire 4C 41.17 system is surrounded by a giant halo of Lyα-emitting gas, which has a diameter of roughly 150 kiloparsec (kpc) *(17)*. We used ALMA's most compact telescope configuration to increase its surface brightness sensitivity. The observing frequency was selected to search for the fine-structure $^3P_1$-$^3P_0$ emission-line of atomic carbon (hereafter [C I]), which has rest-frame wavelength 609 μm (492 GHz). The [C I] transition has an upper energy level (above the ground-state) equivalent to 23.6 K and a critical density $n_{crit} \sim 500$ cm$^{-3}$. It therefore traces molecular gas that has a temperature of 10 to 100 K and volume density of $n_{H2} \leq$ few $\times$ 100 cm$^{-3}$ *(18)*. [C I] can be used as a tracer of molecular $H_2$ gas under a wide range of conditions, including in environments with a high cosmic ray flux *(18, 19)*, or where star formation has not yet commenced *(20)*, although mass estimates rely on assumptions about the gas properties (supplementary text).

The observed distribution of [C I] emission in 4C 41.17 is shown in Fig. S4. By selecting emission at different velocities, we imaged a string of [C I] blobs that align roughly in north-south direction (Fig. 1). Their kinematics gradually transition from higher (north) to lower (south) velocities. The emission extends ~120 kpc north-west of the center of 4C 41.17, which we identify as the radio core at right ascension (R.A.) 06h 50m 52.15s; declination (dec) +41° 30′ 30.8″ (J2000 equinox), using additional observations with the Multi Element Remotely Linked Interferometer Network (MERLIN) telescope (Fig. 1) *(16)*. One end of the [C I] emission is at the approximate location and velocity of 4C 41.17. The [C I] emission forms a coherent structure in both location and velocity, so we co-add the total signal using statistical weighting *(16)*. We find the integrated signal of the emission is inconsistent with the normal distribution of the noise background, with a total significance (neglecting systemic uncertainties) of $5.7\sigma$ (Fig. S2). This statistical calculation *(16)* includes the regions labelled NW1 to NW4 in Fig. 1, but excludes the region W that is inside the Lyα halo.

**Interpretation as a cold gas stream**

We interpret the observed distribution and velocity of the [C I] emission as a stream of cold gas. The stream appears clumpy in Fig. 1, although this appearance is enhanced by the data sampling that we used to image the [C I] in the different regions. The emission extends for at least 100 kpc outside the Lyα halo, although the narrow-band imaging that was used to detect Lyα (Fig.1) was restricted to velocities that do not overlap with those we measure for the [C I] stream *(16)*. The stream is aligned perpendicular to the major axis of both the Lyα halo and the radio jet. It appears to connect to a prominent dark lane in the Lyα emission, roughly 30 kpc west of the central radio galaxy (Fig. 1). We estimate *(16)* that the chance of randomly detecting a spurious stream-like signal of at least $5.7\sigma$ significance in our data is 0.08% ($3.2\sigma$). This calculation assumes that the stream is connected both spatially and kinematically to the [C I] emission in the central region of 4C 41.17, and accounts for confirmation bias (Fig. S2) *(16)*. We used an optimal stacking method to extract the integrated [C I] signal across the entire stream (Fig. 2), from which we calculate the stream's total mass of molecular hydrogen $M_{H2} \sim (6.7 \pm 2.2) \times 10^{10}$ solar masses ($M_\odot$) (supplementary text). The inferred physical properties of the stream are listed in Table S1.

Figure 3B shows a position velocity diagram of the [C I] emission, covering the stream and the central galaxy (Fig. 3A). The projected velocity of the stream steadily decreases at smaller offsets, until it matches the Lyα gas kinematics in region W *(21)*, where the stream connects to the 4C 41.17 system. This is consistent with gas accreting into the Lyα halo. In its outermost region (NW4), the stream has an initial velocity of $v \sim 650 \pm 150$ km s⁻¹ with respect to the gas at the base of the stream (NW1). This is larger than the typical velocities observed for tidal gas tails associated with galaxy interactions *(22)*. The direction and kinematics of the stream appear inconsistent with an outflow, which would likely align along the jet axis and slow down going outward, contrary to the observed [C I] outside the Lyα halo. Stream velocities of ~650 km s⁻¹ at distances of $\gtrsim$100 kpc appear in simulations of gas accretion *(23)* for systems with dark matter halos of total mass $\sim 3 \times 10^{13}$ $M_\odot$, consistent with the value we derive for 4C 41.17 (supplementary text). The width of the stream is spatially unresolved, $\leq$30 kpc, which is not constraining but consistent with simulations *(5)*. The [C I] in the central region of 4C 41.17 is shown in Fig. 4. This [C I] emission is spread across a velocity range of ~1000 km s⁻¹, similar to previous observations of carbon monoxide in the CO($J$=4→3) line *(24)*, where $J$ is the rotational quantum number. Due to the limited spatial resolution of our ALMA data, we cannot study the physical processes that occur in the gas in this central region. Fig. 3B shows additional emission south-east of 4C 41.17, in roughly the opposite direction to the stream (also visible in Figure S4).

**Implications for the assembly of 4C 41.17**

Because the stream contains carbon, it cannot consist of pristine gas from the Big Bang, but must have been enriched with metals (elements heavier than helium). Cosmological simulations predict that low-mass galaxies and giant star-forming clumps flow along accreting gas streams *(2, 13, 25, 26)*, which could enrich the streams with metals due to outflows driven by star formation. Imaging of Lyα emission from cosmic filaments on larger scales has shown that most Lyα emission is powered by ultra-low-luminosity dwarf galaxies *(9)*. Streams and filaments might therefore contain a large number of dwarf galaxies, too faint to be visible in optical imaging. For 4C 41.17, two adjacent star-forming galaxies have previously been detected in near-infrared (*K* band) imaging *(27)*, in a region that shows a deficit in Lyα emission *(17)*. We find that they are close to, but not coincident with, the base of the stream (Fig. 4). Previous studies concluded that these are probably dusty star-forming galaxies which could absorb and scatter the Lyα emission *(17, 27)*. If these galaxies are at the same redshift as 4C 41.17, they might be related to the stream; however, their redshifts have not been measured so we cannot draw further conclusions from them. southeast of these two star-forming galaxies is the prominent dark lane in the Lyα imaging, where the stream appears to connect to 4C 41.17. At this dark lane, Lyα photons might be absorbed by dust or neutral hydrogen gas, potentially provided by the stream.

For the [C I] emission in the central region of 4C 41.17 (Fig. 4), we derive a total mass of molecular hydrogen gas of $M_{H2} \sim (1.4 \pm 0.2) \times 10^{11}$ M$_\odot$ (supplementary text). The bulk of the stars in 4C 41.17 are part of an evolved stellar population, which has been forming stars at a continuously high rate of $\gtrsim$250 M$_\odot$ yr$^{-1}$ for the previous 1 Gyr *(28)*. A recent burst of star formation, possibly enhanced locally by the radio jet *(29)*, has temporarily raised the current star formation rate by an order of magnitude *(29, 30)*. Even a continuous star-formation rate of ~250 M$_\odot$ yr$^{-1}$ would deplete the supply of molecular gas within 4C 41.17 in 560 Myr, so before redshift $z \sim 3$. Assuming the stream is depositing molecular gas into 4C 41.17, we estimate its average accretion rate $\dot{M}_{acc} \sim 450 \pm 180$ M$_\odot$ yr$^{-1}$ (supplementary text). If the stream acquires its gas from a larger reservoir, such as the cosmic web *(9)*, then the stream could potentially supply sufficient gas to 4C 41.17 to sustain the high star formation rate for much longer than the current gas depletion time.

We conclude that the formation of 4C 41.17 is not an isolated process, but linked to the surrounding intergalactic medium.

**Acknowledgments:** The authors thank Guillaume Drouart for useful discussions. This paper makes use of the following ALMA data: ADS/JAO.ALMA#2018.1.01334.S. ALMA is a partnership of ESO (representing its member states), NSF (USA), and NINS (Japan), together with NRC (Canada), MOST and ASIAA (Taiwan), and KASI (Republic of Korea), in cooperation with the Republic of Chile. The Joint ALMA Observatory is operated by ESO, AUI/NRAO, and NAOJ. The National Radio Astronomy Observatory is a facility of the National Science Foundation operated under cooperative agreement by Associated Universities, Inc. MERLIN is a National Facility operated by the University of Manchester at Jodrell Bank Observatory on behalf of STFC. This paper uses observations made with the NASA/ESA Hubble Space Telescope, obtained from the data archive at the Space Telescope Science Institute, which is operated by the Association of Universities for Research in Astronomy, Incorporated, under NASA contract NAS5-26555. **Funding:** Support for Program number HST-AR-16123.001-A was provided through a grant from the STScI under National Aeronautics and Space Administration (NASA) contract NAS5-26555. The MERLIN observations benefited from research funding from the European Community's sixth Framework Programme under RadioNet R113CT 2003 5058187. MDL thanks the Centre National de la Recherche Scientifique / Institut National des Sciences de l'Univers (CNRS/INSU) for their financial and administrative support. NM acknowledges partial support from the Gordon and Betty Moore Foundation through Grant GBMF7392, from the National Science Foundation under Grant No. NSF PHY-1748958, and from the Klauss Tschira Foundation through the Heidelberg Institute for Theoretical Studies (HITS) Yale Program in Astrophysics (HYPA). MVM acknowledges support from grant PID2021-124665NB-I00 by the Spanish Ministry of Science and Innovation (MCIN) / State Agency of Research (AEI) / 10.13039/501100011033 and by the European Regional Development Fund (ERDF) "A way of making Europe". MPT acknowledges financial support through grants CEX2021-001131-S and PID2020-117404GB-C21 funded by MCIN/AEI/10.13039/501100011033. NAH acknowledges support from the Science and Technology Facilities Council (STFC) consolidated grant ST/T000171/1. **Author contributions:**

**Supplementary Materials:**

Materials and Methods

Supplementary text

Figures S1-S5

Table S1

References *(31-93)*

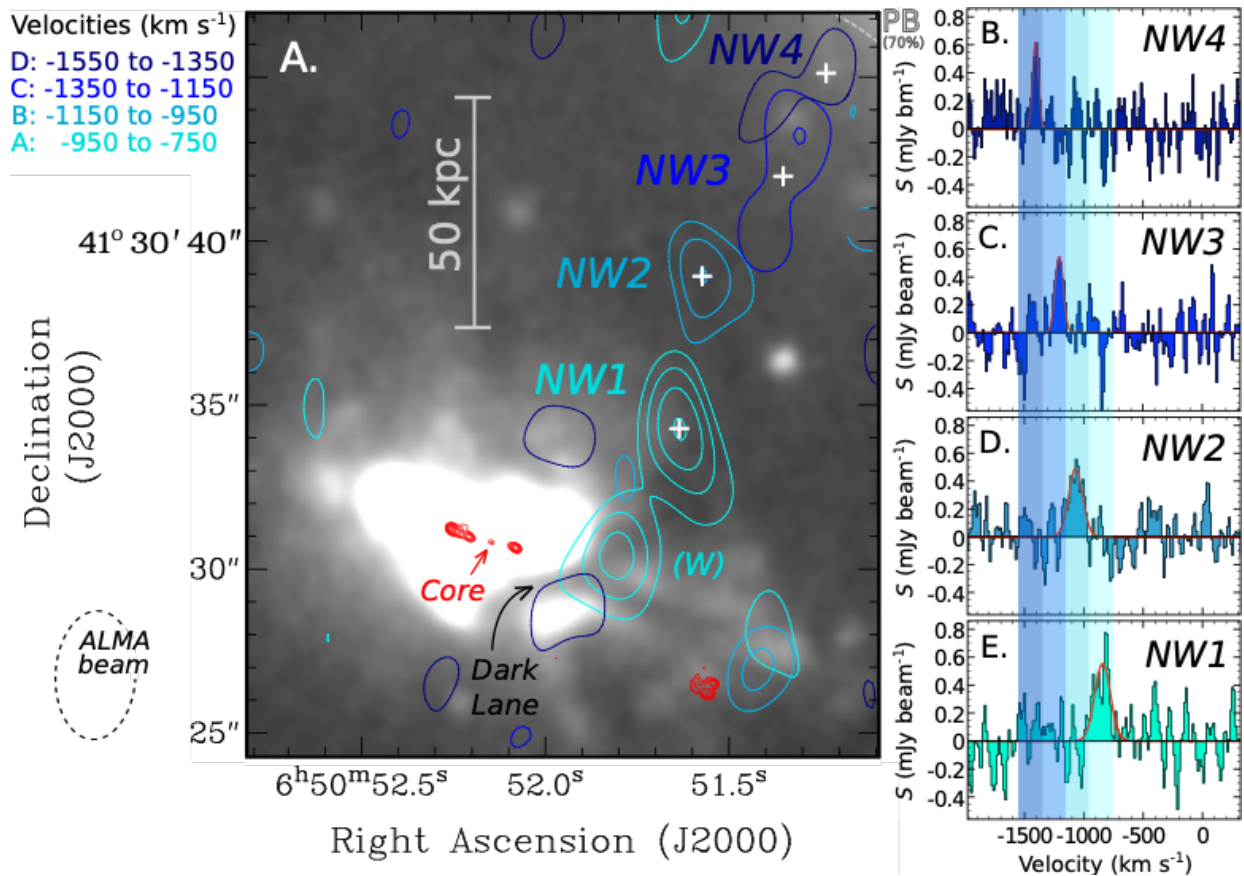

**Fig. 1. [C I] emission around 4C 41.17. (A)** Observed [C I] emission contours, in velocity ranges indicated by the colors in the legend. The background grayscale image shows the Lyα halo *(17)*. Contour levels are at $2\sigma$, $3\sigma$, $4\sigma$, and $5\sigma$, with $\sigma = 0.011$ Jy beam$^{-1}$ km s$^{-1}$ (where Jy is the jansky, 1 Jy = $10^{-26}$ W m$^{-2}$ Hz$^{-1}$). No correction for primary beam (PB) response was applied, so fluxes are progressively attenuated with increasing distance from the center of 4C 41.17. The red contours show the radio source observed with MERLIN *(16)*. Contour levels start at 0.33 mJy beam$^{-1}$ and increase in steps of factor 2. The radio core at the center of 4C 41.17 and the dark lane in the Lyα image are labelled. The scale bar indicates 50 kiloparsecs at the redshift of 4C 41.17. The synthesized ALMA beam is shown by the dashed ellipse in the bottom-left corner. **(B-E)** Spectra of [C I] at the locations north-west of 4C 41.17 marked by white crosses in panel A, labeled NW1, NW2, NW3, and NW4. These spectra were smoothed with a Hanning filter for display. Flux densities $S$ have been corrected for the PB response. The red lines are Gaussian models fitted to each spectrum, and the shaded regions indicate the velocity ranges used in panel A; colors are the same as in the legend. Figure S4 shows [C I] emission at additional velocities, including in the central radio galaxy.

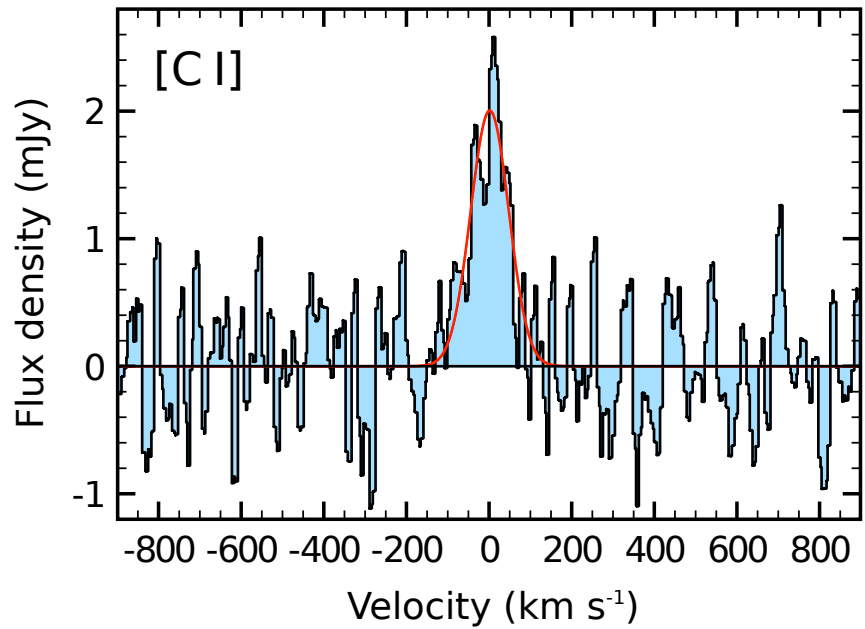

**Fig. 2. Stacked spectrum of the [C I] emission in the stream.** The spectrum includes only [C I]-emitting gas in the stream outside the Lyα halo, i.e. regions NW1, NW2, NW3, and NW4 in Fig. 1. The region labeled W in Fig. 1 (located west of the radio galaxy) was not included in this stack. The independent spectra of the regions NW1 to NW4 were co-added, after shifting the spectra by aligning the velocities of the Gaussian models fitted to the [C I] line in each region (Fig. 1B-E). The stacked spectrum uses channels of 14.5 km s⁻¹ and no Hanning smoothing was applied. The red line shows a Gaussian model fitted to the stacked spectrum, which was used to estimate the physical properties of the stream (see Table S1).

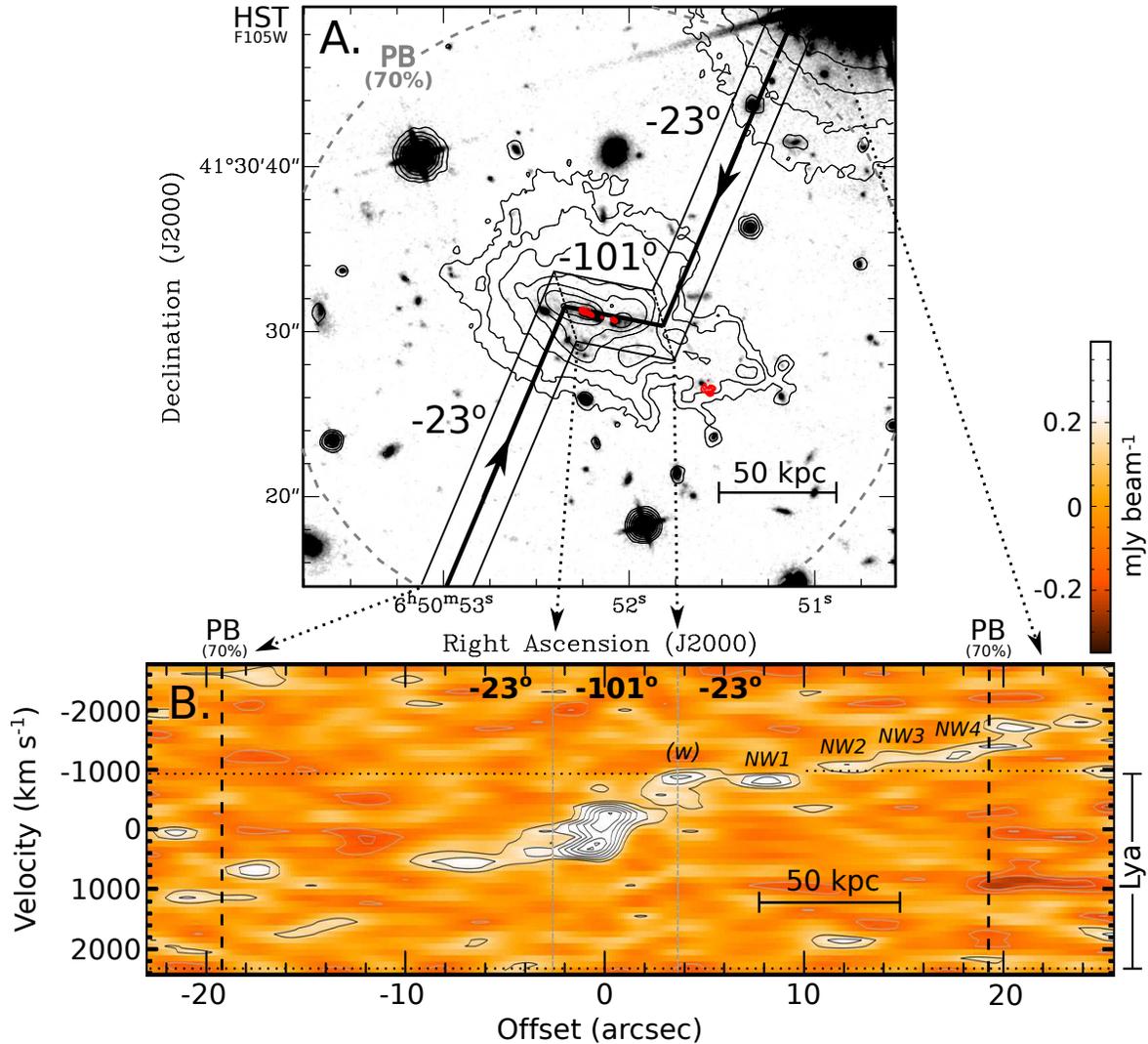

**Fig. 3. Kinematics of the [C I]-emitting gas in the stream.** (**A**) The background is an inverted grayscale image from the Hubble Space Telescope (HST) using the Wide Field Camera 3 (WFC3) and the F105W filter *(30)*, which spans 900 to 1200 nm (188 to 250 nm in the rest-frame at $z = 3.792$). Black contours indicate the Lyα emission from recombining hydrogen gas *(17)*. Red contours show the radio jet as in Fig. 1. The thick black line with arrows shows the direction used to derive the position-velocity diagram (panel B), with the thin black lines marking the boundaries of the included emission (equal to the size of the synthesized ALMA beam). The labeled position angles are measured from north to east. The dashed circle indicates the 70% response level of the primary beam (PB). (**B**) Position-velocity diagram of the [C I] emission in the regions indicated in panel A. Dotted arrows connect equivalent locations in each panel. This assumes the gas has the same redshift as the galaxy, $z = 3.792$. Black contours start at 130 μJy beam$^{-1}$ and increase in steps of 65 μJy beam$^{-1}$; this corresponds to $2\sigma$ with steps of $1\sigma$ at the center of the galaxy (offset of zero). Negative contours are plotted at the same levels in gray. The data have been corrected for the PB response. The vertical dashed lines indicate the 70% response level of the PB, where the noise is 43% higher than at the center of the galaxy. The dotted horizontal lines mark the velocity extent of the Lyα data (panel A and Fig. 1). Regions NW1, NW2, NW3, NW4, and W are the same as in Fig. 1.

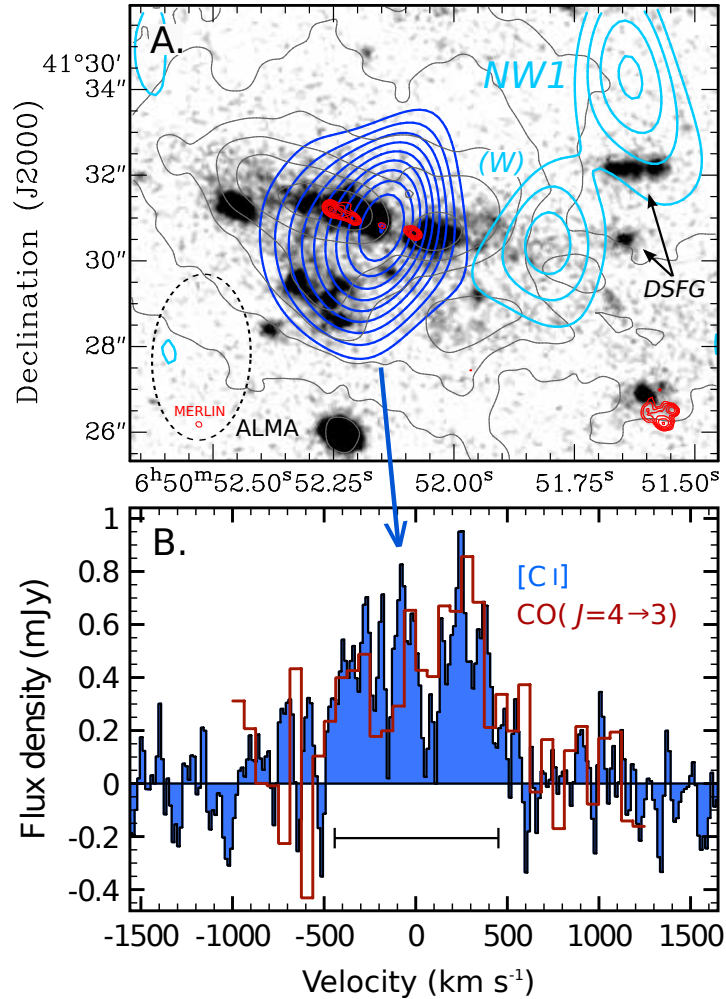

**Fig. 4. Gas in the center of 4C 41.17.** (**A**) Dark-blue contours show the total intensity of [C I] emission in the center of 4C 41.17, overlain on the same HST image as Fig 3A. Contour levels start at $2\sigma$ and increase in steps of $1\sigma$, with $\sigma = 0.038$ Jy bm$^{-1}$ km s$^{-1}$. The light-blue contours show the [C I] emission from part of the stream, as in Fig. 1. The gray contours show the Ly$\alpha$ emission as in Fig. 1, while the red contours show the MERLIN data of the radio jet *(16)*. Black arrows indicate two dusty star-forming galaxies (DSFGs; see also Fig. S4). (**B**) Hanning-smoothed spectrum of the [C I] emission in the center of 4C 41.17. Overlain in dark red is the spectrum of CO($J$=4→3) emission *(24)*, scaled down by a factor of three in flux density for comparison. Both spectra are centered at redshift $z = 3.792$. The black bar indicates the velocity range -440 to +460 km s$^{-1}$ used to integrate the [C I] emission, to obtain the contours in panel A.

# Science

## AAAS

Supplementary Materials for

**A cosmic stream of atomic carbon gas connected to a
massive radio galaxy at redshift 3.8**

Bjorn H. C. Emonts*, Matthew D. Lehnert, Ilsang Yoon, Nir Mandelker,
Montserrat Villar-Martín, George K. Miley, Carlos De Breuck,
Miguel A. Pérez Torres, Nina A. Hatch, Pierre Guillard

*Correspondence to: bemonts@nrao.edu

**This PDF file includes:**

## Materials and Methods

### ALMA observations

We observed 4C 41.17 with the ALMA 12m array for a total on-source integration time of 4.1 hours as part of Cycle 6, project 2018.1.01334.S. Observations were performed on 23 Dec 2018 in the C43-3 configuration for 49 min on-source and during the period 04 to 24 March 2019 using the most compact C43-1 configuration for a total of 196 min on-source. We used ALMA Band 3 with four spectral windows, each covering 1.875 GHz with 2 MHz channels. One of these spectral windows was centered on the redshifted [C I] $^3P_1$-$^3P_0$ line of atomic carbon, with an observing frequency of $v_{obs}$ = 102.705 GHz (rest frequency $v_{rest}$ = 492.16 GHz). The other three spectral windows were targeted on line-free continuum emission. 4C 41.17 was observed at 24 to 26 degrees elevation during all runs, close it its maximum as seen from ALMA. The average levels for the precipitable water vapor (*PWV*) ranged from 0.6 to 3.6 mm across the runs, with an average value of ~2.1 mm. The total bad data flagging, which was dominated by shadowing of antennas, ranged from 27 to 49% across our runs, with an average of 40%. The low elevation decreased the effective length of the baselines in the north-south direction by almost a factor of 2, which increased our surface-brightness sensitivity compared to typical observations at higher elevation.

We used the Common Astronomical Software Applications (CASA) software, version 5.6.1 *(31)*, to calibrate the raw data using the ALMA pipeline v.42866 *(32)*. Subsequently, we manually imaged the pipeline-calibrated data in CASA 5.6.1. To optimize our imaging for low-surface-brightness emission, we discarded baselines in the data taken in C43-3 configuration that were longer than the maximum projected baseline length in the C43-1 data of 100.000× the wavelength (100 kλ), which is ~290nm. We then used a natural weighting to image the [C I] data, resulting in a synthesized beam that gives a spatial resolution of 3.9 × 2.3 arcsec at a position angle of PA = -3 degrees. The uncertainty in the absolute astrometry of these ALMA data is $\delta\theta_{bas}$ ≈ ($\delta\phi_{bas}/2\pi$) $\langle \Theta_{beam} \rangle$ ~ 0.25 arcsec, where $\delta\phi_{bas}$ ~ ($2\pi/\lambda$)($\delta\boldsymbol{B} \cdot \Delta\boldsymbol{k}$) is the phase error in the data, $|\delta\boldsymbol{B}|$ ~ 1 mm the assumed typical error in the baseline-length, $|\Delta\boldsymbol{k}|$ = 10.5° the distance to the phase calibrator B3 0552+398, $\lambda$ = 2.9 mm the wavelength of our observations, and $\Theta_{beam}$ =3.9 arcsec the size of the synthesized beam *(33)*.

We constructed image cubes with two spectral resolutions. One was binned by 5 channels to a resolution of 14.5 km s⁻¹, to show the spectral signature of the [C I] emission. The other was binned by 35 channels to a channel width of 101 km s⁻¹, which was subsequently Hanning smoothed to an effective resolution of 202 km s⁻¹. This velocity resolution corresponds to almost the full width at zero intensity of the stacked [C I] line profile across the stream (Fig. 2), so the latter image cube is optimal for visualizing the spatial distribution of the weak [C I] emission of the stream, and was used for Fig. 3B. For this image cube, we applied a deconvolution and restoration ("cleaning") of signals above a threshold of 0.4 mJy beam⁻¹ channel⁻¹. The two image cubes have root-mean-square (rms) noise levels of 0.24 mJy beam⁻¹ channel⁻¹ for the velocity resolution of 14.5 km s⁻¹, and 0.065 mJy beam⁻¹ channel⁻¹ for the Hanning smoothed 202 km s⁻¹ resolution. The [C I] emission in the center of 4C 41.17 has a systemic redshift of $z$ = 3.792 (Fig. 4) *(34, 35)*. We adopted this redshift to center the velocity in our image cubes.

### MERLIN observations

We observed 4C 41.17 with the Multi Element Remotely Linked Interferometer Network (MERLIN) to image the synchrotron continuum emission from the radio jet at high resolution. We also use the MERLIN data to ascertain the astrometric accuracy of the HST image. We observed 4C 41.17 on 3 June 2005 with MERLIN, which included the antennas at Defford, Cambridge, Knockin, Darnhall, Tabley, and Jodrell Bank (Mark 2 and Lovell dishes). We centered the observations at 1.66 GHz, and used the maximum available bandwidth of 16 MHz. We used the sources 3C 286 for absolute flux density calibration, MRK 0668 for bandpass calibration, and B3 0651+410 for phase-calibration. We observed 4C 41.17 in phase-reference mode, with observations of B3 0651+410 interleaved between each observing scan of 4C 41.17.

We analyzed the correlated data using the Astronomical Image Processing System (AIPS) *(36)*, following standard calibration and imaging procedures. Once calibrated, we imaged the data by applying several iterations of phase- and amplitude self-calibration, which improved the quality of the final image (Fig. S1). The image has a peak flux density of 49.6 mJy beam$^{-1}$ at right ascension 06h 50m 52.26s; declination +41° 30′ 31.2″ (J2000 equinox), and an off-source background rms of 45 μJy beam$^{-1}$, with a synthesized beam of 0.15 × 0.11 arcsec at PA = 52.1 degrees.

## Statistical analysis of the ALMA results

To investigate whether systematic effects could influence our results, we performed a statistical analysis. We binned our data cube to 2.3 × 4.5 arcsec pixels and 202 km s$^{-1}$ channels, and rotated the cube to align the stream in the horizontal direction. The size of the pixels and channels was chosen to match the spatial and spectral resolution shown in Figs. 1 and 3, so each pixel in this re-arranged data cube represents an independent resolution element. The image cube was not corrected for primary beam effects to preserve a uniform rms noise level throughout the cube. Figure S2 shows that this puts the [C I] signal of the stream across regions NW1, NW2, NW3, and NW4 in four adjacent resolution elements of this re-arranged data cube. We refer to these 3D resolution elements as pixels, even though technically the velocity direction traces channels. Because the signal is coherent and found across four adjacent pixels, our method of co-adding the signal is analogous to binning or smoothing the data set. This means that if the co-added signal is statistically significant, then so must be the signal from the individual pixels, even if they individually have a low signal-to-noise. The integrated [C I] signal across these four adjacent resolution elements is detected at a signal-to-noise ratio $SNR = 5.7$. We interpret this as a lower limit, because the optimal stacking performed on the higher resolution data cube resulted in a detection at $SNR = 8.5$ (Fig. 2). Our method of using a re-arranged, low-resolution data cube is therefore not optimized for extracting the maximum signal-to-noise of the [C I] emission in the stream, but it has the advantage that it is a pure representation of the data set. We use this to apply a coherent sampling to the data, which is less affected by confirmation biases than the stacking method, so provides more reliable statistics. Our goal is to estimate the chance that a spurious stream-like signal would be associated with 4C 41.17 at the same $SNR$ as we observe.

To search for features similar to the [C I] stream across this data cube, we first masked the strong line emission from the central radio galaxy by setting the values of the corresponding pixels to zero (Fig. S2). We also masked region W (Fig. 2), and thus excluded it from the statistical analysis, because it lies inside the Lyα halo. We then used the Python programming language with the numpy *(37)*, astropy *(38)*, and matplotlib software packages *(39)* to co-add the signal for all possible combinations of four adjacent resolution elements in the *X-Y* plane (which is the rotated right ascension vs declination plane) in the horizontal, vertical, and two diagonal directions. We did this by stacking the data in velocity in three different ways; by applying a velocity shift of one channel for each consecutive pixel in either the blue- or redshifted direction, or by applying no shift. We left out regions at the edge of the original data cube, where part of the stacked signal would fall outside the primary beam of our observations, but we included the zero-values that we masked out at the location of the radio galaxy in order to study features that stretch across the radio galaxy. The latter has a negligible effect on our statistics. This gave us in total roughly 76,600 different realizations of pixel-combinations with the same dimensions as that of the [C I] stream.

Figure S2B shows the statistical distribution of the signal-to-noise ratio across these 76,600 different realizations. It follows a normal distribution with a standard deviation of $\sigma_{SNR} = 1.0$, with no evidence of systematic effects. We therefore assume that our measured $SNR = 5.7$ for the stream corresponds to a significance level for [C I] emission of $5.7\sigma$. Apart from the emission in the [C I] stream, none of these 76,600 realizations has $SNR \geq 4.2$ (or $\leq -4.2$), as expected for a normal distribution. The only other feature in our statistical analysis that shows an integrated signal that deviates from the noise, at $SNR = 4.6$, is the overlapping combination of regions NW1, NW2, and NW3 (without region NW4), plus an empty pixel from the masked region of the radio galaxy (Fig. S2A). This demonstrates that no other features that mimic the signal of the stream, either with positive or negative flux, are present in our image cube down to 75% of the flux density of the stream ($SNR > 4.2$).

For the total [C I] emission significance of $5.7\sigma$, the probability ($P$) of obtaining a sample with a standard normal random variable ($z_n$) that is more extreme than our observed $SNR$ of the stream-like signal is $P(z_n > 5.7) = 6.0 \times 10^{-9}$, based on a one-sample right-tailed Z-test. The chance that one such spurious signal appears among the 76,600 different realizations of pixel-combinations that we used to perform our statistics is therefore about 0.046%. However, this is the chance that a spurious signal with $SNR \geq 5.7$ and the same dimensions as the [C I] signal of the stream appears anywhere within the data cube when searching blindly. This is not the most appropriate statistical analysis for testing any possible arrangement that would be interpreted as a stream. The observed stream aligns both spatially and kinematically with the bright [C I] emission in the central region of 4C 41.17, so a blind search will under-estimate the statistical significance of the signal. Simultaneously, the statistical significance of the signal is over-estimated as a result of confirmation biases.

To estimate the true significance of the stream, we consider the following aspects:

i). Constraints and biases in the selection of the signal within our re-binned image cube, including (a) the location, (b) the direction, and (c) the length of the stream.

ii). Biases in the binning of our original image cube. Because the original image cube over-sampled the signal in right ascension and declination, this binning was necessary to obtain a re-binned image cube that consists of independent resolution elements, making it suitable for a statistical analysis. However, this binning may have severely limited the degrees of freedom for the possible orientation of the stream, which we need to take into account as a separate confirmation bias.

This is the only ALMA data set observed with this specific instrument setup and integration time, hence our sample size is one. This therefore does not affect our statistics.

Given the uniform sampling of our re-binned image cube, we can obtain an estimate of the factors that affect the signal selection in our re-binned image cube:

i.a). Position of the stream: we find that the stream connects spatially and kinematically to the bright [C I] emission in the central region. In our re-binned image cube, there are 148 3D pixels adjacent to the central [C I] emission where region NW1 could have been located to connect to 4C 41.17. We consider this an upper limit, because we do not take into consideration that the stream ends at the location of a prominent dust-lane, which would reduce the number of possible pixels where we would expect the stream to connect to the central reservoir by an order of magnitude.

i.b). Direction of the stream: we assume that the stream is aligned away from the central [C I] emission, without further constraining the alignment between the stream and the central gas reservoir. In our re-binned image cube, each starting pixel for region NW1 has 26 adjacent pixels. Taking into consideration the overlap between signals, and the fact that the stream is aligned away from the central gas reservoir, roughly 13 of those pixels could have hosted region NW2, and therefore mark the direction of the stream. We also consider this as an upper limit, because the stream extends perpendicular to the direction of the [C I] and radio source in the central galaxy, but we do not require this stricter alignment.

i.c). Length of the stream: the stream has a length of 4 pixels in our re-binned image cube. However, if we consider that the stream could be as short as 3 pixels (~90 kpc), which is similar to the radius of the Lyα halo, or as long as 5 pixels (~160 kpc), which would stretch the stream to the half-power width of the primary beam, then this would raise the number of possible stream signals by a factor of 3.

Therefore, the number of pixel combinations in our re-binned image cube that can provide a signal that is spatially and kinematically aligned with the bright central [C I] emission and ranges from 90-160 kpc in length is roughly $5.8 \times 10^3$. This results in a probability of $3.5 \times 10^{-5}$ that our re-binned image cube would host a spurious stream-like signal that connects to the central gas reservoir in 4C 41.17 and has at least the same SNR as the observed stream.

Next, we consider the confirmation bias related to the binning of our original image cube. This includes any bias in the position and direction of the stream that was not considered by the much coarser sampling of our re-binned image cube. Figure S3 shows how the *SNR* of the signal changes when applying a shift in pixels or rotation angle during the re-sampling of the original image cube. The *X*-direction of the re-binned image cube is the direction perpendicular to the stream. We binned the original image cube by 15 pixels in this direction. Figure S3A shows that if we shift our sampling in the *X*-direction by more than $|\Delta X_{pix}| = 3$ pixels in either direction from the peak *SNR*, then the probability increases that the stream signal that we detected was spurious. Given that we binned by 15 pixels in *X*-direction, we therefore multiply the number of possible pixel combinations for the stream in our re-binned image cube by a factor $\frac{15}{2|\Delta X_{pix}|+1} = 2.1$ to compensate for the confirmation bias in binning our original image cube in *X*-direction. In the same way, in the *Y*-direction along the stream, we binned our original image cube by 30 pixels. Figure S3B shows that if we shift our sampling in the *Y*-direction by more than $|\Delta Y_{pix}| = 6.5$ pixels in either direction from the peak *SNR*, then the probability increases that the stream signal that we detected was spurious. We therefore multiply the number of pixel combinations in our re-binned image cube by a factor $\frac{30}{2|\Delta Y_{pix}|+1} = 2.1$ to compensate for the confirmation bias in binning the image cube in *Y*-direction. This also an upper limit, because it does not take into account that the velocity of the stream also gradually changes in Y-direction.

Our statistical analysis is also biased by the fact that we rotated our original image cube in the *X-Y* plane to align the stream in vertical direction, and that we only sampled our re-binned image cube at intervals of 45 degrees in rotation angle, assuming square pixels. Figure S3C shows that if we change the rotation of our original image cube by a rotation angle ($\Delta R_{deg}$) more than $|\Delta R_{deg}| = 8$ degrees in either direction from the peak *SNR*, then the probability increases that the stream signal that we detected was spurious. We therefore multiply the number of pixel combinations in our re-binned image cube by a factor $\frac{45}{2|\Delta R \deg|+1} = 2.6$ to compensate for the confirmation bias from rotating our original image cube. Finally, in the velocity direction, we binned our original image cube by two channels. Because the original image cube was Hanning smoothed, these two channels were not mutually independent. We therefore multiply the number of pixel combinations in our re-binned image cube by a factor 2.0 to compensate for our confirmation bias in binning the image cube in velocity direction. In total, we therefore estimate that the overall bias from rotating and binning the original image cube results in a factor of $2.1 \times 2.1 \times 2.6 \times 2.0 = 23$ larger number of possible orientations for the stream than that we estimated for the re-binned image cube on which we performed our statistics. This means that the previously calculated probability of $3.5 \times 10^{-5}$ should be increased by this factor of 23. Therefore, our estimate for the overall chance that our re-binned image cube would host a spurious stream-like signal with at least the same *SNR* as the [CI] stream is $8 \times 10^{-4}$ (0.08% or 3.2$\sigma$). In other words, we expect that 1 in 1250 data cubes with the same properties as ours would randomly show a spurious stream-like signal with *SNR* $\geq 5.7$.

We regard this as a conservative estimate, because *(a)* a more optimal stacking of the signal revealed a significantly higher *SNR* = 8.5 for the stream signal (Fig. 2), *(b)* the observed stream is aligned perpendicular to the radio axis and connects to a prominent dark lane in the Ly$\alpha$ image, but we did not include those constraints in our statistical analysis, *(c)* we did not take into account the [C I] emission in region W, where the stream connects to the central gas reservoir (Fig. S4).

**Supplementary Text**

Properties of 4C 41.17
4C 41.17 is a high-redshift radio galaxy *(14, 15, 40)*. Figure 4 shows 4C 41.17 consists of a group of galaxies and clumps that are likely in the process of merging *(14, 15, 27)*, as expected from the hierarchical process of galaxy formation. 4C 41.17 is surrounded by a giant halo of Ly$\alpha$-emitting gas (Fig. 1), which has a diameter of about 150 kpc *(17, 41)*. Extended X-ray emission is also present in the halo *(42)*. This system will likely evolve into a single giant cluster galaxy *(40, 43)*, so we refer to the entire region surrounded by the giant Ly$\alpha$ halo as 4C 41.17. 4C 41.17 has a total stellar mass $M_* = 4.2 \times 10^{11}$ M$_\odot$ *(44)*, and infrared (8 to 1000 $\mu$m rest-frame) luminosity $L_{IR} \sim 2 \times 10^{13}$ L$_\odot$ from star formation alone *(45)*. 4C 41.17 also contains large amounts of dust, detected in sub-millimeter emission *(46-49)*.

To derive physical properties of 4C 41.17, we assume a cosmology with Hubble constant $H_0 = 71$ km s$^{-1}$ Mpc$^{-1}$, ratio between the matter density and critical density of the universe $\Omega_M = 0.3$, and ratio between the energy density and critical density of the universe $\Omega_\Lambda = 0.7$. At a redshift of $z = 3.792$, this puts 4C 41.17 at a luminosity distance of 33,166 megaparsec (Mpc) and 1 arcsec corresponds to 7.0 kpc in angular scale *(50)*.

Radio source
The galaxy in the center of 4C 41.17 hosts a supermassive black hole that produced a jet, which emits synchrotron radiation *(14, 51)*. The radio source aligns with ultra-violet (UV) continuum and emission-line gas, indicating that the propagating radio jet has triggered star formation *(29, 35, 52)*. Our MERLIN image (Fig. S1) shows the double-lobed structure of the radio source seen in previous work *(14, 51)*, with a two-sided jet across the inner ~15 kpc and a hot-spot ~60 kpc to the south-west. In the center, the MERLIN image has a faint blob of emission between the inner lobes, which aligns with the brightest emission in the rest-frame UV HST image. We interpret this as most likely the core of 4C 41.17. Based on the alignment of the core in the MERLIN and archival HST data, we infer that that the relative astrometry between the optical and MERLIN data is accurate to about 0.1 arcsec, which is smaller than the previously estimated astrometric accuracy of the ALMA data of ~0.25 arcsec *(16)*.

The [C I] stream is aligned perpendicular to the direction of propagation of the radio source, which is also the major axis of the Lyα halo and most likely the ionization cone of the active galactic nucleus (AGN) (Fig. S4). This perpendicular alignment makes it unlikely that the [C I] emission in the stream is the result of photo-ionization by the AGN, as has been inferred from observations of a [C I] emitter in the halo of another radio galaxy at redshift 3.6 *(53)*, or by an outflow driven by the radio source, given that those are typically well aligned with the jet axis *(54)*.

Molecular gas mass

Molecular gas predominantly consists of $H_2$, but this molecule is difficult to observe directly. A tracer for molecular gas is carbon, which has sub-millimeter emission from its singly ionized [C II] and neutral [C I] forms, or from molecular carbon-monoxide (CO). While millimeter observations often use CO as a tracer for molecular gas, [C I] emission can also be a tracer of the molecular $H_2$ content of cold gas under a wide range of conditions, with potentially brighter emission than CO from regions with high gas turbulence or a high cosmic ray flux *(18-20)*.

Mass estimates from both [C I] and CO rely on assumptions about physical conditions of the gas, such as excitation, opacity, metallicity, and abundances. Previous studies found that [C I] and CO line emission are similar in distribution and fairly uniform in intensity ratio inside molecular clouds in the Milky Way *(55, 56)*. For molecular cloud complexes in starforming galaxies, the intensity ratio of CO and [C I] varies over scales of a few 100 parsec, likely due to gradients in the gas excitation or opacity *(57)*. This makes [C I] a less reliable as mass tracer for studies that resolve galaxy disks *(57)*. However, when considering the integrated [C I] and CO luminosities across a large sample of galaxies from redshift 0 to 6, mass conversion factors are adequate for deriving reliable mass estimates from the ground-transition of either [C I] or CO, i.e., [C I] $^3P_1$-$^3P_0$ or CO($J$=1→0) *(58)*. Atomic carbon is on average highly subthermal, so mass estimates from the higher-level [C I] $^3P_2$-$^3P_1$ transition are much less constrained *(59)*.

To derive molecular gas masses from our ALMA data of [C I] $^3P_1$-$^3P_0$, we first estimate the mass of [C I] ($M_{[C I]}$) from the [C I] luminosity ($L'_{[C I]}$) listed in Table S1, using *(60)*:

$$M_{[C I]} = 5.706 \times 10^{-4} \, Q(T_{ex}) \, \frac{1}{3} \, e^{23.6/T_{ex}} \, L'_{[C I]} \quad (60),$$

with $Q(T_{ex}) = 1 + 3e^{-T_1/T_{ex}} + 5e^{-T_2/T_{ex}}$ the [C I] partition function. $T_1 = 23.6$ K and $T_2 = 62.5$ K are the energies above the [C I] ground-state, while $T_{ex}$ is the excitation temperature, which we assume to be $T_{ex} = 30$ K *(60)*. To convert this to the molecular gas mass, we assume that the neutral carbon abundance ($X_{[C I]}$) relative to the abundance of $H_2$ ($X_{H_2}$) is similar to the value found in the Milky Way Galaxy:

$$X_{[C I]}/X_{H_2} = M_{[C I]}/(6 \, M_{H_2}) = 2.2 \times 10^{-5} \quad (61),$$

This is also close to the values found for molecular gas in the circum-galactic medium of the Spiderweb Galaxy, where $X_{[C I]}/X_{H2} = 1.5 \times 10^{-5}$ *(62)*. For high-redshift star-forming galaxies, typical values range from $(2\text{-}5) \times 10^{-5}$ *(63, 64)*. We estimate the molecular gas mass in the stream as $M_{H2} \sim (6.7 \pm 2.2) \times 10^{10}$ $M_\odot$ (Table S1). For the uncertainty, we added in quadrature the uncertainty in the measured integrated intensity (Table S1) and the 30% differences between the $X_{[C I]}/X_{H2}$ value observed for the Milky Way *(61)* and that for the CGM of the Spiderweb Galaxy *(62)*. This estimate is consistent to within 5% compared to mass estimates based on assumptions presented in other studies *(58, 59)*.

We regard our estimate for the molecular gas mass as a lower limit, because we did not apply corrections for metallicity or effects of the cosmic microwave background (CMB). Simulations predict that the [C I]-to-$H_2$ conversion factor increases at lower metallicity, following roughly $\alpha_{[C I]} = M_{H2} / L'_{[C I]} \propto (Z/Z_\odot)^{-1}$ *(20, 65)*, with $M_{H2}$ the molecular gas mass, $L'_{[C I]}$ the [C I] luminosity, and $Z/Z_\odot$ the metallicity relative to the solar value. Previous optical studies that trace the CGM of massive galaxies at $z \sim 2$ to 3 along lines of sight towards background quasars found metallicities ranging from the solar value to 1% of the solar value *(66,67)*, with the latter value attributed to a stream *(67)*. However, because the metallicity of the gas in our stream is unknown, we did not apply this correction factor, which would increase our mass estimate. We also did not apply a correction for effects of the CMB, which simulations predict could decrease the observed line intensity by about 10% (40%) of the intrinsic value for gas at 50K (20K) at that redshift *(68, 69)*, thus increasing the mass estimate. The unknown gas temperature, and lack of observational confirmation of the effect of the CMB on line emission, would make any CMB correction uncertain.

In the central ~40 kpc of 4C 41.17, we detect [C I] emission with a velocity-integrated flux of $\int_v S_{[C I]} \, \delta v = (0.47 \pm 0.08)$ Jy km s$^{-1}$ (Fig. 4), where $S_{[C I]}$ is the [C I] flux density and the integral is taken over the full velocity width of the emission line. This flux corresponds to a [C I] line luminosity of $L'_{[C I]} = (1.4 \pm 0.3) \times 10^{10}$ K km s$^{-1}$ pc$^2$ *(70)*. If we adopt the Milky Way conversion factor $X_{[C I]}/X_{H2} = 2.2 \times 10^{-5}$ *(61)*, and furthermore adopt the same assumptions as above ($T_{ex} = 30$ K, $Z/Z_\odot = 1$, and no CMB correction), then this implies a molecular gas mass of $M_{H2} = (1.4 \pm 0.2) \times 10^{11}$ $M_\odot$ for the central gas reservoir in 4C 41.17.

Mass accretion rate from the stream

We estimate the average mass accretion of molecular gas deposited by the stream onto 4C 41.17 as $\dot{M}_{acc} = (M_{H2\text{-}stream} \times v_{stream})/l_{stream} \approx 450 \pm 180$ $M_\odot$ yr$^{-1}$, with $M_{H2\text{-}stream} \sim (6.7 \pm 2.2) \times 10^{10}$ $M_\odot$ the mass of molecular gas in the stream, $l_{stream} \sim 100$ kpc the length of the stream, and $v_{stream} \sim 650 \pm 150$ km s$^{-1}$ the initial velocity of the stream. This initial velocity is the difference between the gas velocities in the outermost region NW4 and the region NW1, where the stream connects spatially and in velocity to the gas in the Lyα halo. This is thus the same as the velocity gradient averaged over the length of stream. The velocity difference between the gas in region NW4 and the systemic velocity of the central radio galaxy ($v_{sys\text{-}RG}$) is larger (~1400 km s$^{-1}$), but the Lyα-emitting gas in region W has a velocity of about -750 km s$^{-1}$ with respect to $v_{sys\text{-}RG}$ *(21)*, which is consistent with the [C I] velocity at the base of the stream. In estimating the average mass accretion rate, we only consider the [C I] detected across the regions NW1, NW2, NW3, and NW4. In other words, we only consider the molecular gas that lies outside the observed Lyα halo. The stream in Figs. 1, 3, and S4 is clumpy, which could cause fluctuations in the mass accretion rate.

Nevertheless, over the ~150 Myr crossing time of the stream, the average mass accretion rate of molecular gas that we observe is roughly $\dot{M}_{acc} \sim 450 \pm 180$ M$_\odot$ yr$^{-1}$. This value is based on the properties of the stream as projected in the plane of the sky, without accounting for the unknown inclination.

Star formation rate of 4C 41.17

The average molecular mass-accretion rate of ~450 M$_\odot$ yr$^{-1}$ is a factor of two times higher than the long-lived star formation rate (SFR) of 4C 41.17, $SFR \sim 250$ M$_\odot$ yr$^{-1}$ (28). The galaxy likely contains an evolved stellar population with a phase of star formation that peaked about 700 Myr earlier, with a more recent burst of star formation that peaked about 30 Myr ago (28). The value $SFR \sim 250$ M$_\odot$ yr$^{-1}$ is the current star formation rate within the evolved stellar population. We compare this value with the accretion rate of the stream, because the crossing time of the stream is roughly 150 Myr, which is a factor of five longer than the age of the recent starburst. The recent starburst has raised the current star formation rate temporarily by an order of magnitude, giving 4C 41.17 its current infrared brightness. However, this phase is likely short lived compared to the longer timescale of the gas accretion provided by the stream, and probably the result of jet-triggered star formation that is localized along the radio jet (29, 30, 35, 52), or an instantaneous event like a major galaxy merger (28) or accretion of a clump carried by the stream (13, 25). Other studies, which assumed that all of the far infrared emission is the result of an instantaneous burst of star formation, found a higher star formation rate of $SFR \sim 3300$ M$_\odot$ yr$^{-1}$ (45). This is consistent with sub-millimeter observations of the thermal dust in 4C 41.17 (46), which imply a $SFR \sim$ 2000 - 3000 M$_\odot$ yr$^{-1}$. Spectro-polarimetric observations of the rest-frame UV emission imply a star formation rate between 200 - 1600 M$_\odot$ yr$^{-1}$ when corrected for extinction and our assumed cosmology (29). We therefore regard $SFR \sim 250$ M$_\odot$ yr$^{-1}$ as a lower limit, which could easily be several times higher. Our gas accretion rate of $\dot{M}_{acc} = 450 \pm 180$ M$_\odot$ yr$^{-1}$ is also a lower limit, because we only consider the molecular gas phase in this single stream.

A SFR of ~250 M$_\odot$ yr$^{-1}$ and molecular gas reservoir of $M_{H2} \sim 1.4 \times 10^{11}$ M$_\odot$ leads to a depletion timescale for the current reservoir in 4C 41.17 of ~560 Myr. This is similar to values derived for high-redshift star-forming galaxies (63); it could be substantially shorter if the SFR is higher. This implies that the current gas reservoir would be consumed by $z$=2.9. Accretion of additional gas into 4C 41.17 is required to sustain the SFR over a longer timescale.

Comparison to simulations of stream-fed accretion

Simulations of galaxy evolution that include a cold accretion mode, in which a large fraction of the gas accretion penetrates into the dark matter halo as filaments of cold gas (1, 2, 71), predict that the accretion process occurs through an average of three streams. However, the dominant stream can carry up to two thirds of the total gas accretion rate (72), implying that the other two streams would appear several times fainter in observations. If this applies to [C I] emission, it would make any additional streams undetectable, or only marginally detectable, in our ALMA observations. Our observations of additional emission south-east of 4C 41.17 could perhaps indicate a second stream.

Simulations of cold-mode accretion mostly do not trace cold gas below the Ly$\alpha$-emitting limit of ~$10^4$ K. Simulations that do include gas cooling below $10^4$ K have shown that galaxy disks that are fed by a steady supply of cold streams fragment into giant clumps that host a large fraction of the star formation in the disks (13). The HST imaging of 4C 41.17 shows no evidence of star formation along the stream before it reaches 4C 41.17. Simulations of heating and dissipation within gas streams themselves, using idealized, non-cosmological simulations that mimic the conditions in the CGM of massive high-redshift galaxies (73-75), show that 10 to 50% of the gravitational energy of the inflowing gas can be dissipated into radiation, with some of it converted to Lyα emission (74). As simulated streams interact with the CGM, they can also entrain gas from the environment through Kelvin-Helmholtz and thermal instabilities (75).

We compare these simulations with the observed properties of the [C I] stream. First, we investigate whether the properties of the stream as it falls into the dark-matter halo of 4C 41.17 are as expected based on the system's virial mass. The 4C 41.17 system has a stellar mass $M_* \sim 4.2 \times 10^{11}$ $M_\odot$ (44). Simulations of dark-matter halos predict that at $z \sim 4$ the total stellar mass of a system is roughly 1% of the halo mass (76), which suggests that the estimated virial mass of 4C 41.17 is roughly $M_{vir} \sim 4 \times 10^{13}$ $M_\odot$. An independent estimate of $M_{vir}$ can be obtained from previously published long-slit spectroscopy of the Lyα emission along the major axis of the Lyα halo (21), which showed kinematics consistent with rotation in the outer parts of the halo, with a rotational velocity of at least $v_{rot} \sim 400$ km s$^{-1}$ at a radius of $\sim 70$ kpc. The virial mass associated with such a rotating structure is expected to be at least $M_{vir} \sim 3 \times 10^{13}$ $M_\odot$ (77). The virial radius $R_{vir}$ for a virial mass of $M_{vir} \sim 3 \times 10^{13}$ $M_\odot$ is $R_{vir} \sim 190$ kpc (78). From the virial theorem $M_{vir} = R_{vir} \times v_{vir}^2 / G$, where G the gravitational constant and $v_{vir}$ is the virial velocity, we estimate $v_{vir} \sim 8 \times 10^2$ km s$^{-1}$. Hydrodynamical simulations predict that the infall velocity of a stream ($v_{stream}$) is related to $v_{vir}$ by $v_{stream} = \eta \times v_{vir}$, where $\eta \sim 0.9$ for massive dark-matter halos (23). We therefore expect the initial velocity of the stream $v_{steam, init} \sim 700$ km s$^{-1}$. This is consistent with the initial stream velocity of $v_{stream} \sim (650 \pm 150)$ km s$^{-1}$ that we previously derived for the gas in outermost region NW4.

Simulations predict that a stream at $z \sim 4$ feeding a halo of $M_{vir} \sim 3 \times 10^{13}$ $M_\odot$ has a width $\sim 10$ kpc, with a range of $\sim 2$ to 40 kpc (75), which is consistent with our observed upper limit of $\leq 30$ kpc on the width of the stream. We set an upper limit on the average gas turbulence within the stream from the full width at half the maximum intensity (FWHM) of the [C I] line profile, $FWHM_{[C I]} \sim 110$ km s$^{-1}$, measured from the stacked spectrum of the stream (Fig. 2). This is consistent with simulations that predict the turbulent velocity of gas in a cold accreting stream to be only about 10 to 30% of the initial velocity of the stream (74).

In the simulations, as the streams accelerate towards the center of the halo, they become denser and narrower. Simulations find a typical hydrogen volume density of $n_H \sim 0.03$ to 1 cm$^{-3}$ at the virial radius, but this is before accounting for any perturbations or turbulence in the stream, and assuming the stream has a temperature of $T \sim 10^4$ K. If the pressure stays constant, then $n_H \propto T^{-1}$, so if we assume that the temperature in the stream drops by a factor of 100 (from $10^4$ to $\sim 100$ K), the mean density of the cold gas would be $n_H \sim 10$ to 100 cm$^{-3}$. As the streams flow towards the center, the density increases with galactocentric distance roughly as $(r/R_{vir})^{-2}$ (25, 75). It is therefore plausible for the gas in the streams within $\sim 100$ kpc of the central galaxy to reach $n_H \sim 500$ cm$^{-3}$, the critical density of [C I]-emitting gas. As discussed in the main text, low-mass galaxies or giant star-forming clumps flowing along accreting streams (2, 13, 23, 25, 26) could explain the carbon content of the stream in 4C 41.17.

Simulations predict the mass accretion rate for a virial mass of $M_{vir} \sim 3 \times 10^{13}$ $M_\odot$ at $z \sim 4$ is $\dot{M}_{acc} \sim (1$ to 10$) \times 10^3$ $M_\odot$ yr$^{-1}$ for a single stream, assuming a baryon fraction of 17% (25, 75). This is higher than the overall mass accretion rate of $\dot{M}_{acc} \sim 450 \pm 180$ $M_\odot$ yr$^{-1}$ that we derived from [C I] in 4C 41.17, possibly because our results only include [C I]-emitting gas, or our previous assumption of solar-value for the gas metallicity was too high.

The cold stream with its average velocity width of $FWHM_{[C\ I]} \sim 110$ km s$^{-1}$ also resembles filamentary structures of molecular gas observed in and around Brightest Cluster Galaxies (BCGs) at low redshifts *(79-84)*. Both observations *(81-83)* and simulations *(85-87)* have shown that many filaments of molecular gas in low-redshift BCGs are related to gas cooling in the updraft of X-ray bubbles produced by radio jets. We consider this scenario unlikely for 4C 41.17, because the stream is oriented orthogonal to the radio jet and extends much further out than the radio continuum.

## Molecular gas in the center of 4C 41.17

The [C I] emission in the center of 4C 41.17 is concentrated in the region around the radio galaxy (Fig. S5). While the emission peaks on top of the central radio galaxy, it also stretches into the CGM between the radio galaxy and its nearest companion galaxies (Fig. S5). At the lowest velocities (Fig. S5A-C), the [C I] peaks at the location of an inner dark lane seen previously in the Lyα image, with some emission stretching in the direction of the outer dark lane near the base of the stream *(17, 24)*. At the highest velocities, the [C I] stretches south-east of the radio galaxy. This [C I] emission at the highest and lowest velocities peaks in regions with diffuse rest-frame UV emission, between the brightest peaks in the HST image, covering a scale of ≲ 20 kpc. This is similar to the molecular CGM observed in CO($J$=1→0) at the centers of two other giant Lyα halos at $z\sim2$, the Spiderweb Galaxy *(88)* and Mammoth-I *(89)*, albeit on a smaller scale. For the Spiderweb Galaxy, the diffuse rest-frame UV emission is known to trace in-situ star formation *(90)* that is fueled by the widespread molecular gas in the CGM *(62, 88)*.

We observe twice as much [C I] flux as a previous study using the Northern Extended Millimeter Array (NOEMA) telescope of the Institut de Radioastronomie Millimétrique (IRAM) *(35)*. The bright [C I] emission at the central velocities in our ALMA data is located along the optical morphology of the radio galaxy, as found in the NOEMA data. However, the ALMA observations have more short baselines and a synthesized beam twice as large as the NOEMA data, making them more sensitive to low-surface-brightness emission from widespread [C I]-emitting gas. Compared to NOEMA, we recover additional [C I] flux in the ALMA data from widespread molecular gas, particularly at negative velocities where the emission in 4C 41.17 mostly peaks between the galaxies. NOEMA also has a smaller primary beam than ALMA, reducing its sensitivity at large distances from the central galaxy, such as in the stream.

The central distribution of the [C I] is similar to the distribution of molecular gas previously seen in CO($J$=4→3) with the IRAM Plateau de Bure Interferometer (PdBI) *(24)*. Although the PdBI observations did not cover the velocity range of the stream, our [C I] spectrum matches the previously published CO($J$=4→3) spectrum at the location of the radio galaxy (Fig. 4). The distribution of the CO($J$=4→3) emission showed a main CO component aligned with the radio core and inner dark lane seen in the Lyα imaging, and a second CO component co-spatial with an outer dark lane in the Lyα emission towards the SW (Fig. S5) *(24)*. The bright [C I] emission in the central region does not appear to extend as far as the outer dark lane in Fig. S5, but at lower velocities the [C I] in the stream connects to the central gas reservoir in this region (Fig. S4), consistent with the CO($J$=4→3) results.

Previous observations of 4C 41.17 searched for CO($J$=1→0), but resulted in a non-detection *(91, 92)*. The upper limit was high enough to conclude that the bulk of the H$_2$ gas in the central region of 4C 41.17 must have an excitation similar to molecular gas in low-$z$ starburst galaxies *(92)*.

**Supplementary Figures, and Tables**

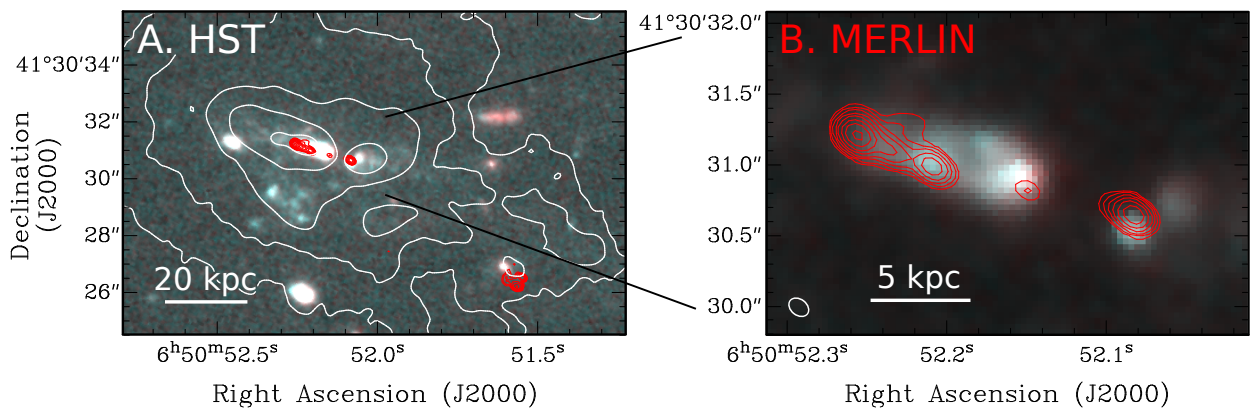

**Fig. S1. Radio galaxy 4C 41.17.** (A) Two-color image taken with the Hubble Space Telescope Wide Field Camera 3 (WFC3) in the F105W (blue color) and F160W (red color) filters *(30)*, with overlain contours of the Lyα halo in white *(17)* and the radio synchrotron emission observed with MERLIN in red. Lyα contours start at 9% of the peak intensity of the halo and increase by factors of 3. Contour levels of the MERLIN data start at 0.33 mJy beam$^{-1}$ and increase by factors of 2. (B) zoom-in of the central region of the radio galaxy. The weak central radio component co-spatial with the bright HST emission is probably the core of 4C 41.17. The synthesized beam is shown as the white ellipse on the bottom-left. We estimate that the astrometry of the HST image matches that of the radio data within ~0.1 arcsec.

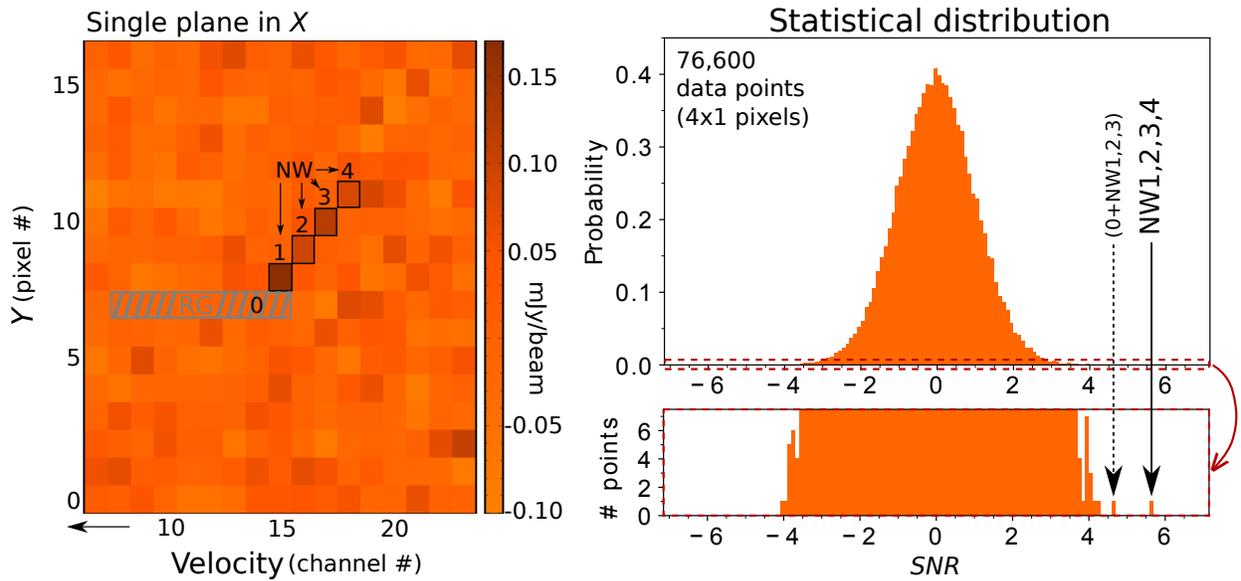

**Figure S2. Statistical significance test for the [C I] stream.** (A) a section of the re-binned and rotated [C I] image cube showing the emission from the stream in regions NW1, NW2, NW3, and NW4 across four adjacent pixels in vertical (*Y*) and velocity-direction, while aligned in the same plane in horizontal (*X*) direction. For the pixels that cover the radio galaxy (RG), the flux density was set to zero (gray hatched region). (B) Statistical distribution of the signal-to-noise ratio (*SNR*) of features in the image cube with the same morphological properties as the [C I] stream in terms of length and width. For this test, we stacked four consecutive pixels in horizontal, vertical and both diagonal directions in the *X-Y* plane, with a velocity shift of -1, 0, and 1 channels. Taking the signal from all these stacked elements resulted in 76,600 data points. (C) Zoom in of the red dashed box shown in panel B. On the y-axis, instead of the probability, the true number of data points (from the total of 76,600 used in the statistics) are shown. The black arrows in both panel B and C highlight the *SNR* value of the co-added regions NW1 + NW2 + NW3 + NW4 and 0 + NW1 + NW2 + NW3 from panel A. The stream detection across regions NW1, NW2, NW3, and NW4 is 5.7$\sigma$ above the Gaussian noise.

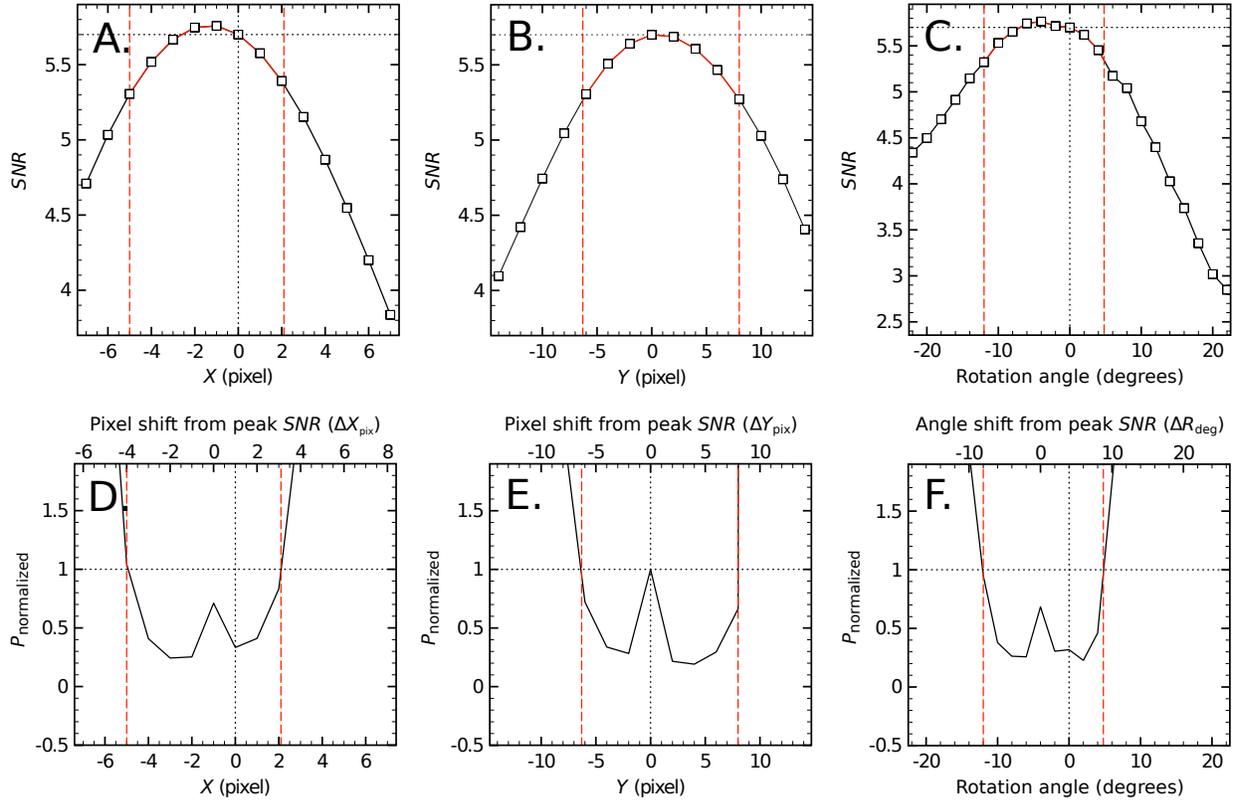

**Figure S3. Confirmation bias tests from re-sampling of the original image cube.** Each panel shows the effect on the statistical significance of the stream from rotating and binning the original image cube in the *X-Y* plane, for which the original pixels were not mutually independent. Our original image cube was first rotated and then binned by 15 pixels in the *X*-direction, 30 pixels in the *Y*-direction, and 2 pixels in the velocity direction to produce a re-binned image cube, which consists of independent resolution elements, on which we performed our statistics (Fig. S2). The plots show how a shift during the binning process in (a) *X*-direction, (b) *Y*-direction, and (c) *X-Y* rotation angle affects the SNR of the stream signal (panels A-C) and the overall probability that the stream-signal is spurious (panels D-F). (A-C) The horizontal axis represents the shift in pixels when binning our original data cube, with 0 (dotted vertical black line) being the image cube in Fig. S2. The vertical axis shows the *SNR* of the stream after re-binning, with the dotted horizontal black line being the SNR=5.7 derived from our statistical analysis. (D-F) Normalized probability, $P_{normalized} = P(z_n > SNR)/((|2\Delta_{pix}| + 1) \times P(z_n > 5.7))$ of obtaining a sample with a standard normal random variable $z_n$ that is more extreme than the measured *SNR* of the stream-like signal compared to the probability of a signal with *SNR* = 5.7, corrected for a factor $|2\Delta_{pix}|+1$ to reflect the decreasing degrees of freedom, where $\Delta_{pix}$ stands for $\Delta X_{pix}$ (panel D), $\Delta Y_{pix}$ (panel E), or $\Delta R_{deg}$ (panel F). This correction assumes that the drop in SNR is roughly symmetric for positive and negative pixels around the peak *SNR*. Because the pixels in the original image cube are not mutually independent in right ascension and declination, only those degrees of freedom where $P(z_n > SNR)$ drops more than the factor $|2\Delta_{pix}|+1$ are relevant for the statistical analysis. Values for $P_{normalized} \leq 1$, the region between the vertical dashed red lines, indicate that the corresponding shift in pixels while re-binning the original image cube does not increase the overall chance that the stream is spurious.

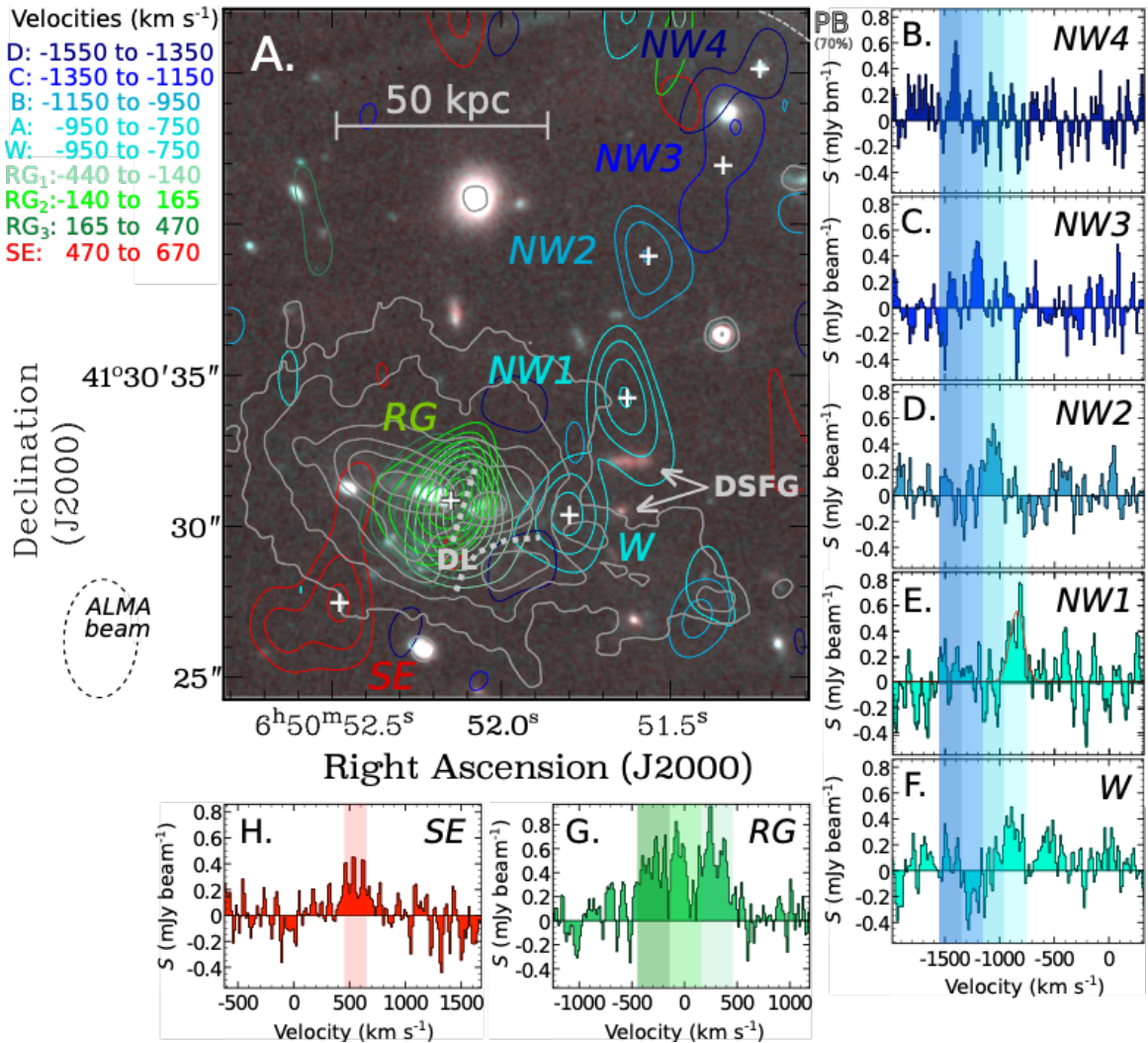

**Fig. S4. All [C I] emission around 4C 41.17.** (A) The blue and cyan [C I] contours are the same as Figure 1A; the green (marked as radio galaxy `RG') and red (marked as south-east `SE') contours show additional velocity channels. The background image is the same as Fig. S1A. The [C I] emission is integrated across the velocity ranges indicated in the legend. Blue and red contour levels start at $2\sigma$ and increase by $1\sigma$, with $\sigma = 0.011$ Jy bm$^{-1}$ × km s$^{-1}$. The green contours start at $3\sigma$ and increase by $1\sigma$, with $\sigma = 0.015$ Jy bm$^{-1}$ × km s$^{-1}$. No correction for primary beam (PB) response was applied. The gray contours show the Lyα halo (background image in Fig. 1A). Two red galaxies in the HST imaging at the base of the stream are probably dusty star-forming galaxies (DSFGs) *(27)*, marked with arrows. The dotted gray lines indicate two prominent dark lanes (DL) previously identified in the Lyα imaging *(17,24)*. Other symbols and labels are the same as Figure 1. (B-H) Spectra of [C I] extracted at the locations of the white crosses in panel A, which capture regions NW1, NW2, NW3, and NW4 of the stream, as well as the radio galaxy (RG), the region west of the radio galaxy (W), and the region south-east of the radio galaxy (SE). The spectra are Hanning smoothed to a velocity resolution of 29 km s$^{-1}$ and values have been corrected for the PB response to give flux densities. The colored shaded regions correspond to the velocity ranges in the legend.

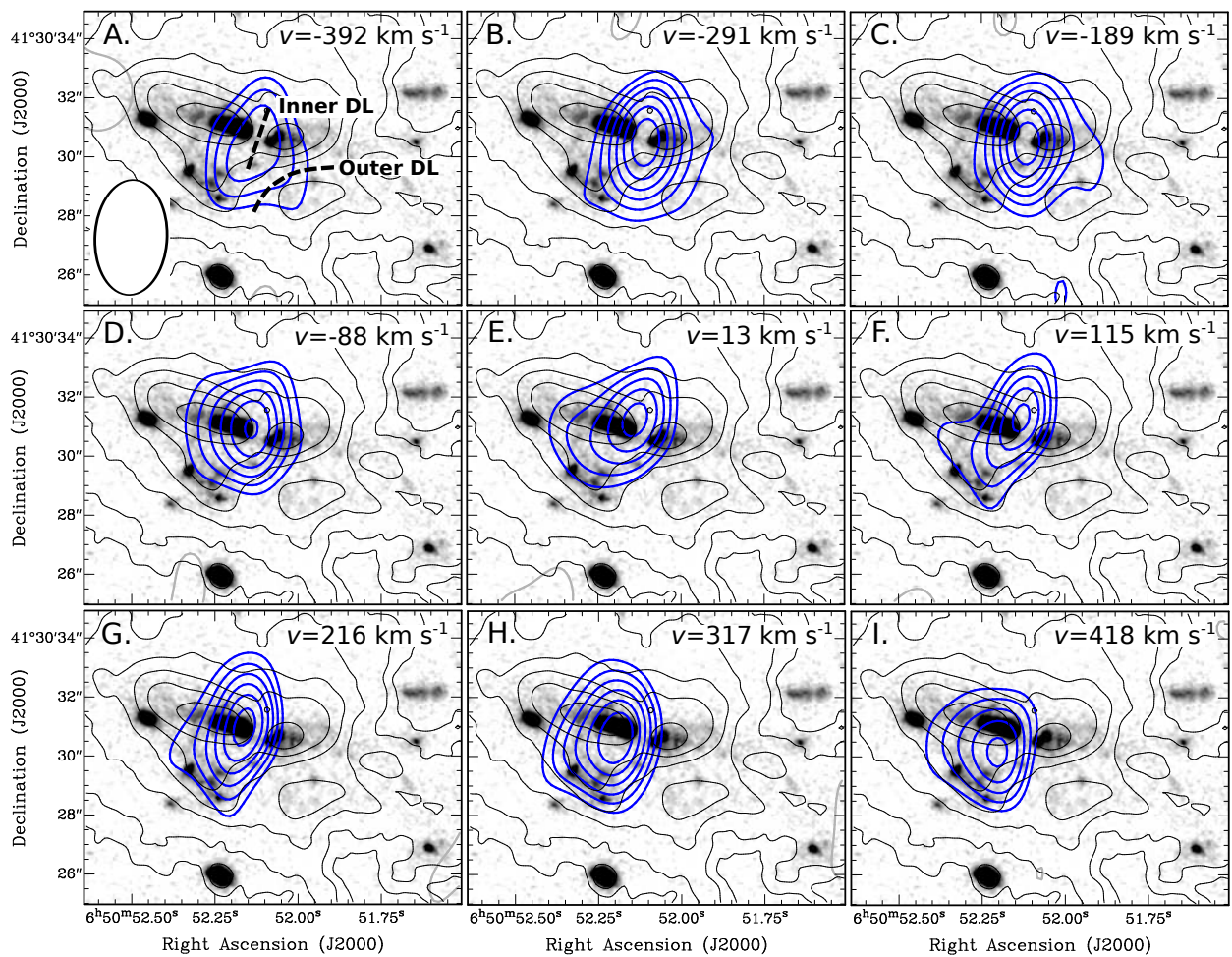

**Fig. S5. Channel maps of [C I] emission in the central region of 4C 41.17.** Blue contours indicate the [C I] emission at different velocities, labelled in each panel. Contour levels start at $2.5\sigma$ and increase in steps of $1\sigma$, with $\sigma = 0.065$ mJy bm$^{-1}$. The black contours show a zoom-in of the Ly$\alpha$ emission from Fig. S4. The background is the HST F105W image from Fig. 3. The ALMA beam is shown by the black ellipse in the bottom-left corner of panel A, while the dashed lines in panel A indicate the location of two prominent dark lanes (DL) visible in Ly$\alpha$ imaging *(17, 24)*.

**Table S1. Derived properties of the stream.**

| Quantity | Symbol | Value & unit |
|---|---|---|
| Length[a] | $l_{stream}$ | ~100 kpc |
| Width[b] | $w_{stream}$ | ≤30 kpc |
| Velocity gradient[c] | $\Delta v_{stream}$ | $650 \pm 150$ km s$^{-1}$ |
| Velocity width[d] | $FWHM_{[C\ I]}$ | $110 \pm 20$ km s$^{-1}$ |
| Integrated flux[d,e] | $\int_v S_{[C\ I]}\,\delta v$ | $(0.23 \pm 0.03)$ Jy km s$^{-1}$ |
| Luminosity[d,f] | $L'_{[C\ I]}$ | $(7.1 \pm 0.9) \times 10^9$ K km s$^{-1}$ pc$^2$ |
| Mass [C I][g] | $M_{[C\ I]}$ | $(8.9 \pm 1.1) \times 10^6$ M$_\odot$ |
| Mass H$_2$[g] | $M_{H2}$ | $(6.7 \pm 2.2) \times 10^{10}$ M$_\odot$ |

a.    The observed length of the stream in the NW direction, taken to be the [C I] emission at angular offsets from about +6″ to +20″ in Fig. 3. The true length of the stream could be larger, because primary beam effects limit our sensitivity for tracing the stream beyond 120 kpc from the center of 4C 41.17.

b.    The stream width is spatially unresolved, so this upper limit corresponds to the width of the synthesized beam.

c.    The [C I] velocity width, velocity-integrated flux, and luminosity of the stream are derived from the Gaussian model fitted to the stacked spectrum in Figure 2. *FWHM* is the full width at half the maximum intensity of the [C I] line. $S_{[C\ I]}$ is the [C I] flux density.

d.    Velocity gradient averaged over the full length of the stream.

e.    Uncertainty in the integrated intensity includes uncertainty in the Gaussian fitting of the stacked spectrum (Fig. 2), uncertainty due to the noise in the spectrum across the full width of the line *(93)*, and a 10% systematic uncertainty in the absolute flux calibration of the ALMA data. We have not applied a correction for the effects of the cosmic microwave background *(68, 69)*.

f.    Calculated following *(70)*.

g.    See Supplementary text.